\documentclass[fleqn,usenatbib]{mnras}

\usepackage{newtxtext,newtxmath}
\usepackage[T1]{fontenc}
\usepackage{ae,aecompl}
\usepackage{graphicx}
\usepackage{amsmath}
\usepackage{amssymb}
\usepackage[normalem]{ulem}

\newcommand{\caps}[1]{{\scshape{#1}}}
\def\apj{ApJ}
\def\aj{AJ}
\def\mnras{MNRAS}
\def\acta{Acta Astron.}
\def\pasp{PASP}

\def\nat{Nature}

\def\an{Astronomische Nachrichten}

\def\dw{DW~Cnc~}

\title[New Observations of \dw]{New observations of \dw: where is the 38 min signal?}

\author[Segura et al.]{
O. Segura Montero,$^{1}$\thanks{e-mail:  osegura@astro.unam.mx}
S. H. Ram\'irez,$^{1}$
and
J. Echevarr\'ia,$^{1}$
\\
$^{1}$Instituto de Astronom\'ia, Universidad Nacional Aut\'onoma de M\'exico, Ciudad Universitaria 04510, CDMX, Mexico\\
}

\date{Accepted XXX. Received YYY; in original form ZZZ}

\pubyear{2020}

\begin{document}
\label{firstpage}
\pagerange{\pageref{firstpage}--\pageref{lastpage}}
\maketitle

\begin{abstract}
We present extensive radial-velocity observations of the intermediate polar \dw during its 2018-2019 low state. We show that the 86 min signal, associated with the orbital period is strong in our radial velocity analysis, power spectrum search, as well as in our Doppler Tomography. However, we find that the velocity modulation associated with the 70-min beat period and the 38-min spin cycle are dramatically weaker than previously observed. We put forward two interpretations for this change. The first is that a sudden drop into a low state detected in 2018-2019 caused an episode of low mass transfer from the companion, thus inhibiting the light-house effect produced by the rebound emission. The second is that this is a consequence of a rare outburst detected in 2007 by \citet{cea08}. We find this post-outburst hypothesis to be less likely. If the first scenario is correct, we predict that \dw\ will recover its intermediate polar characteristics. A new ephemeris is presented by combining \citet{pea04} radial velocities with ours. 

\end{abstract}


\section{Introduction}
\label{intro}
Cataclysmic Variables (CVs) are interacting binaries, which consist of a compact primary star, and a late--type secondary (a slightly evolved main sequence star filling its Roche Lobe) which transfers matter to the usually more massive white dwarf (WD). An accretion disk is formed around the primary star, and the material leaving  the inner Lagrangian point, collides somewhere along the already formed disk producing a bright spot.\citet{wan71} and \citet{sma71} establish this classical model which works rather well for most dwarf novae, novalike variables, and old novae. However, if the magnetic field of the WD is very strong ($\geq$ MG), then no accretion disk is formed and the material falls directly onto the magnetic poles. The accretion energy is reprocessed into an emission beacon which is observed with a periodicity equal to the orbital period, as the magnetic field lines lock the rotation of the compact star with the orbital period. These synchronously rotating binaries are called polars. Alternately, if the magnetic field is not strong enough (0.1~-~10 MG), an external disk can be formed, but the inner regions are inhibited by the Alfven radius. Therefore some of the accreting material  will form an external disk, while the innermost material will be channelled through the lines of the magnetic field. In this case, the rotation of the WD (usually known as the spin rate or spin period) will not be synchronized with the orbital period. If an external ring is present, this will produce not only a lighthouse effect with a spin period, but also a beat period with the external disk. These are the so called intermediate polars. A comprehensive review may be found in \citet {war95}. 

\dw is a short period intermediate polar. It was identified by \citet{ste81} as a variable star with the Byurakan 40 inch Schmidt-camera, possibly a U Gem star, and designated as \dw by \citet{kea82}. The object was classified also as a Dwarf Nova by \citet{kea88}, based on its spectrum at low resolution, which shows strong Balmer lines, weak He I lines and a rather weak He II $\lambda$ 4686 \AA\ emission line (about 10~times weaker than H$\beta$). A visual magnitude range of 15 - 17.5 was roughly estimated by \citet{ste81}, while infrared magnitudes were derived from the 2MASS Second Incremental Data Release: J~=~14.66; H~=~14.32 and K = 14.01 \citep{hea02}. From photometry spanning 61 days (actually 40 observed nights), \citet{uea02} found the object in a high-state with $R_c = 14.68 \pm 0.07$ mag and proposed that the object was not a Dwarf Nova, but rather a non-magnetic nova-like system, with strong quasi periodic oscillations with center periods at 37.5~$\pm$~0.1 and 73.4~$\pm$~0.4 min. They proposed two interpretations for \dw: (i) a permanent super-humper below the period minimum of hydrogen-rich cataclysmic variable, or (ii) a nova-like variable having an orbital period over 3 hours. 

\citet{rea04} (hereinafter Rea04) presented the first time-resolved spectroscopic study of \dw and determined an orbital period of 86.10 $\pm$ 0.05 min. They found that the radial velocity of the Hydrogen and He I lines are also modulated with a second period of 38.58 $\pm$ 0.02 min, a value also supported by their time-resolved photometry. They proposed that \dw is a magnetic VY~Scl (see \citet{hes00} and \citet{rea12}, for recent reviews on VY Scl systems), and also that \dw\ is an intermediate polar below the period gap,  as they tentatively associate the photometric and spectroscopic 38-min signals with the white dwarf spin period. \citet{pea04}~(hereinafter~Pea04) presented an almost simultaneous photometric and spectroscopic study. In particular their spectroscopic radial velocity analysis yielded strong detections at two periods, 86.1015(3) min and 38.58377(6) min, which they interpreted as the orbital period $P_{orb}$ of the binary, and the spin period $P_{spin}$ of a magnetic white dwarf, respectively. They found an additional strong signal at 69.9133(10) min, which coincides with the {\it beat period}. As they point out, these periods are the landmark of the "DQ Her-type" or intermediate polar class of cataclysmic variables. They also found evidence for a weak periodic signal
at 110.85 minutes. As they put it: {\it this does not seem related to any of the other clocks in the binary, and we leave it as an unsolved problem}. \citet{tls08} derived a distance of 257(+79, -52) pc, from parallax estimations, which is in agreement with the measurement obtained by Gaia DR2 of 4.7900 $\pm$ 0.0572 mas or about 209 pc. 

\citet{cea08} reported the first detection of a $\sim$ 4 mag outburst, reaching V~$\sim$~11.36 on January 25, 2007. This is, so far, the only outburst detection of DW~Cnc. Their immediate follow up observations on January 29, 2007  still show a strong signal at 38.6 min. And from  observations up to February 4, 2007, they also find another strong signal at a different frequency of the beat-period, at 73.73  min, similar to the result of \citet{uea02}.

\citet{nea19} report positive {\it XMM-Newton} observations in 2012 in the range 0.3 - 10 Kev. From their light curves, they are able to confirm the existence of a period at 37.7~$\pm$~4.5 min, and find from the OM light curve a signature for a period at 75 $\pm$ 21 min, both consistent within the errors to the spin and beat-periods. 

\begin{figure*}
	\includegraphics[width=2\columnwidth]{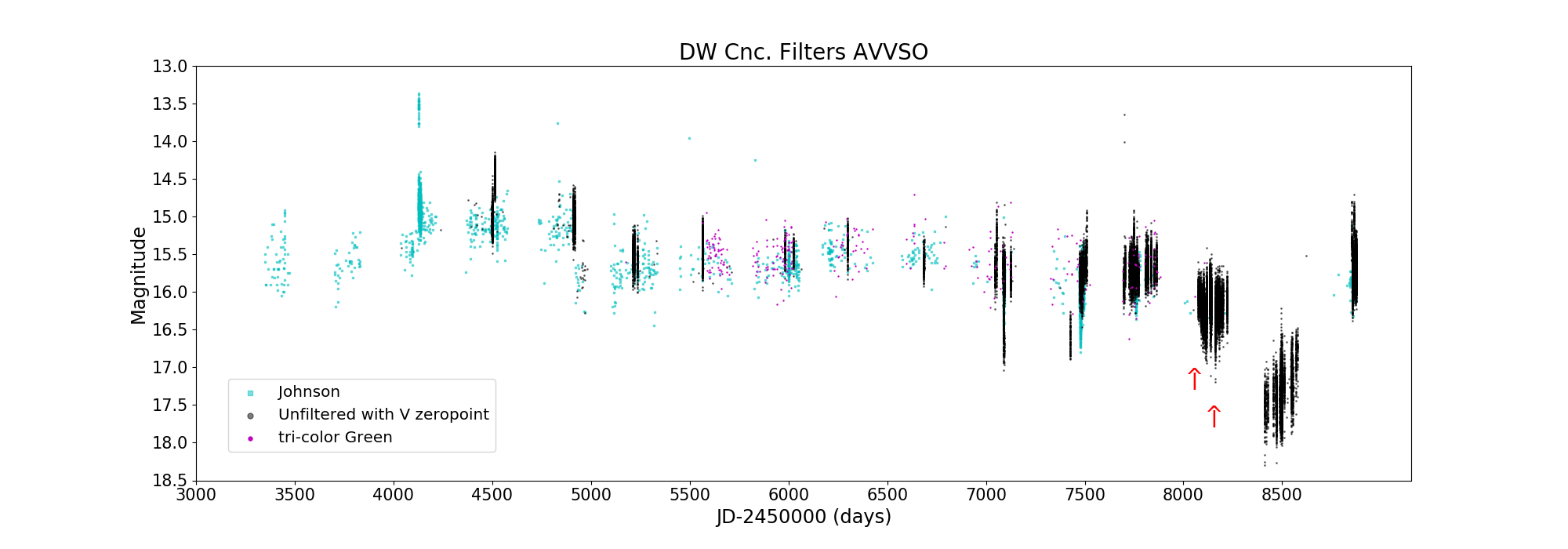}
    \caption{AAVSO Light curve from almost 20 year of observations. The red line represent the time of our spectral observations.}
    \label{aavso}
\end{figure*}

As \dw is claimed mainly by 
Rea04 to be a VY Scl system, we present in Figure~\ref{aavso} the {\sc aavso}  data, collected for almost 20 years. This classification will be discussed in Section~\ref{sec:discussion}.

In this paper we present a new spectroscopic study of DW~Cnc. In Section~\ref{sec:orbper} we derive the radial velocity curves from the H$\alpha$ and He I $\lambda$ 5876 \AA  ~lines, and obtain their semi-amplitude $K_1$. In Section~\ref{power-spec} we discuss the power spectrum signals based on our calculated radial velocities. In Section~\ref{ephem} we present new ephemerides of the system and in Section~\ref{tomo} we use doppler tomography to help us understand the nature of this intermediate polar. Finally in Section~\ref{sec:discussion} we present a discussion on the dramatic change of the different signals discussed in this paper, and present a couple of possible solutions to understand and to solve this change.

\section{Spectroscopic Observations}
\label{sec:obs}
\begin{figure}
  \includegraphics[width=1\columnwidth]{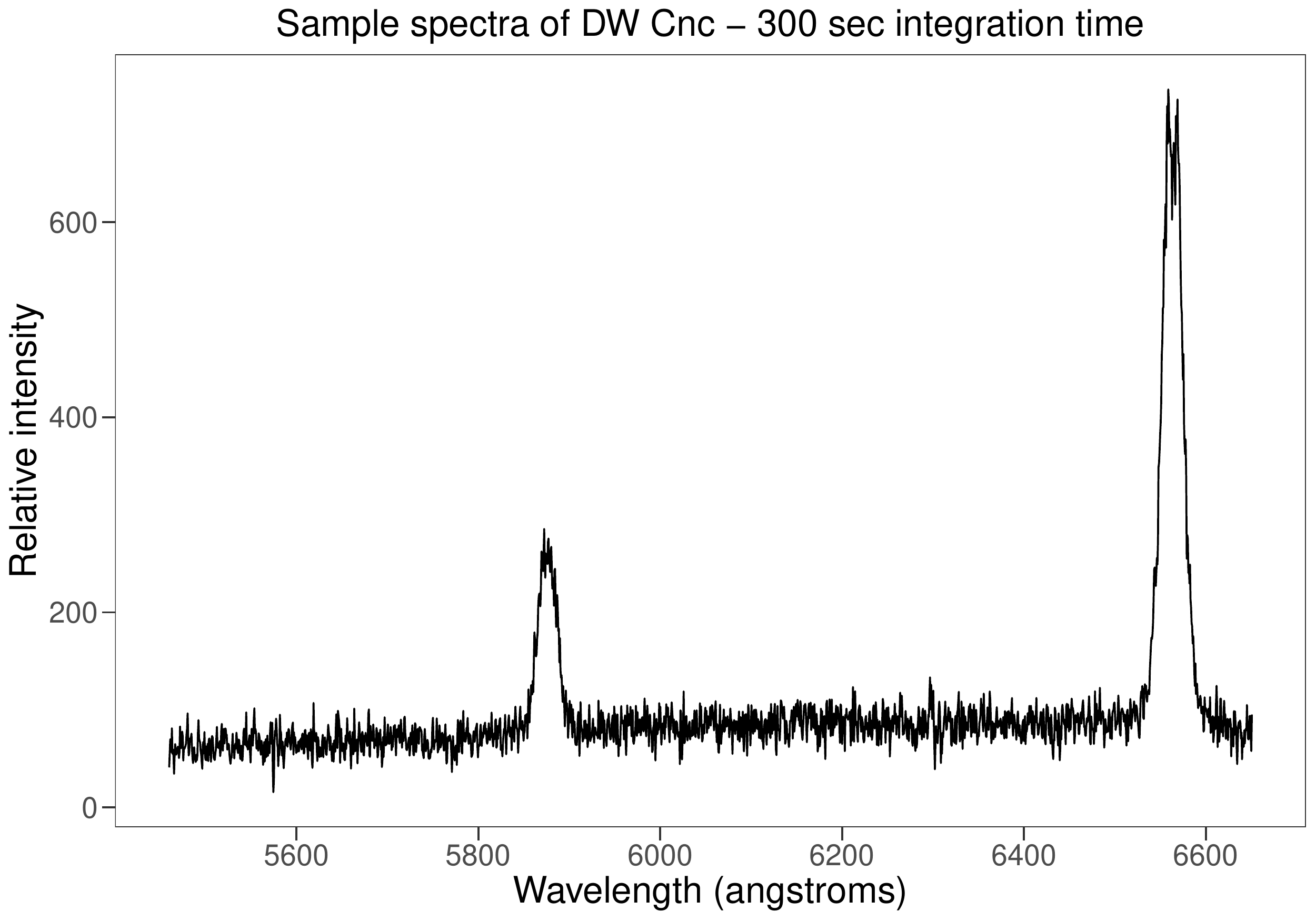}
    \caption{A sample spectrum of the nearly 450 spectra obtained in this work.  The H$\alpha$ and the Helium line $\lambda$ 5876 \AA\ are often double peaked (see text).}
    \label{fig:spectrum}
\end{figure}

 Spectra were obtained with the 2.1m telescope of the Observatorio Astron\'omico Nacional at San Pedro M\'artir, using the Boller and Chivens spectrograph and a e2V CCD detector in the 5400 - 6600~ \AA ~range (resolution $\sim$ 1200), on the nights of 2017 Oct 12, and 2018 January 19 and 21--25. The exposure time for each spectrum was 300~s. More than 450 spectra were obtained\footnote{We will put a link here to access the full radial velocities like: Available at the CDS via anonymous ftp to...}. A single spectrum is shown in Figure~\ref{fig:spectrum} as an example. The H$\alpha$ line is strong, and at times appears double peaked. There is also a strong HeI~$\lambda$~5876~\AA~line, which also shows a complex profile. Standard~{\sc iraf}\footnote{ IRAF is distributed by the National Optical Astronomy Observatories, which are operated by the Association of Universities for Research in Astronomy, Inc., under cooperative agreement with the National Science Foundation.}
procedures were used to reduce the data.   The log of observations is shown in Table~\ref{tab:speclog}.
 
 \begin{table}
\centering
	\fontsize{10}{15}\selectfont 
	\caption{Log of observations for the October 2017 and January 2018 spectroscopy.}
    \label{tab:speclog}
    \begin{tabular}{lccccccccc}
       \hline \hline
       \noalign{\smallskip}
       
Spectroscopy & Julian Date     & No of   & Exposure \\
Date         & (JD - 2450000)  & Spectra & Time     \\
\hline
12 Oct  & 8038 &  18 & 300 s \\
19 Jan  & 8137 &  27 & 300 s \\
21 Jan  & 8139 &  78 & 300 s \\
22 Jan  & 8140 &  95 & 300 s \\
23 Jan  & 8141 &  69 & 300 s \\
24 Jan  & 8142 &  86 & 300 s \\
25 Jan  & 8143 &  92 & 300 s \\
\hline
       \noalign{\smallskip}
\end{tabular}
\end{table}

\section{Radial velocities}
\label{sec:orbper}

We have fitted the radial velocities of H$\alpha$ and HeI $\lambda$ 5876 \AA\ emission lines with a circular orbit of the form 

\begin{equation}
V(t) = \gamma + K \sin\left(2\pi\frac{t - t_0}{P_{orb}}\right),
\end{equation}

where $V(t)$ are the measured radial velocities of the individual spectra, $\gamma$ is the systemic velocity, $K$ is the radial velocity semi-amplitude, $t_0$ is the time of inferior conjunction of the donor, and $P_{orb}$ is the orbital period.

An orbital period of 86.10169~$\pm$~0.00031 min was found using a power spectrum analysis. This is described in Section~\ref{power-spec}. In our analysis, we found no substantial difference by including the October 2016 data. Therefore, to keep a compact data set, we have calculated the radial velocity results for the 2017 data only. We set the orbital period as a fixed parameter, and derive $K$, $\gamma$ and $t_0$ as free parameters.


To measure the radial velocites from the emission lines we used a {\sc convrv} task within the {\sc rvsao} package in {\sc iraf}. This task was given to us privately by J. Thorstensen in 2008, who describes it as a task that computes the velocity of a line for a set of spectra by convolving the line with an antisymmetric function and taking the line centre to be the zero of this convolution, based on the algorithms described by \citet{say80} and \citet{sha83}. This task works in two ways. With the {\sc dgau} option the function convolved is the derivative of a gaussian. In this case the result is interpreted as the FWHM of the line you are fitting. In the second option ({\sc gau2}) the convolution function consists of a positive gaussian and a negative gaussian with a given separation and width.

\begin{figure}
	\includegraphics[width=0.9\columnwidth]{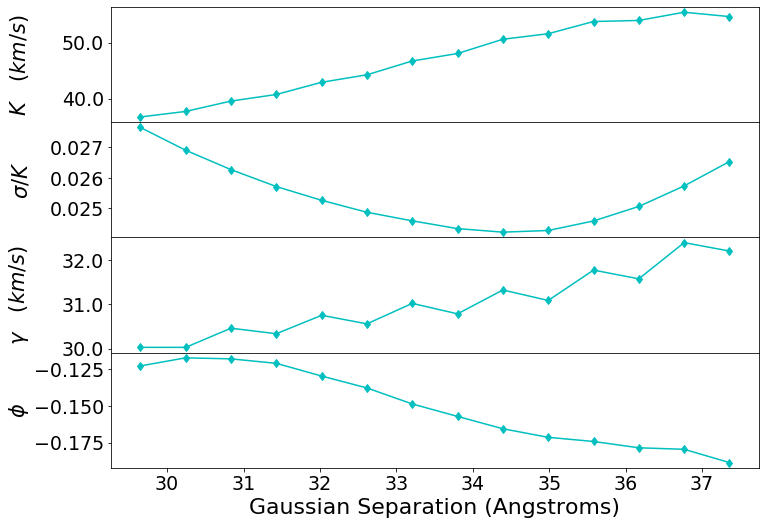}
    \caption{Diagnostic Diagram of the radial velocity curve using the wings of the H$\alpha$ emission line. The best solution is for a Gaussian separation of 34.6 \AA ~ (58 pixels) and a width of 3.5 \AA ~ (6 pixels).}
    \label{fig:diag}
\end{figure}

\begin{figure}
	\includegraphics[width=\columnwidth]{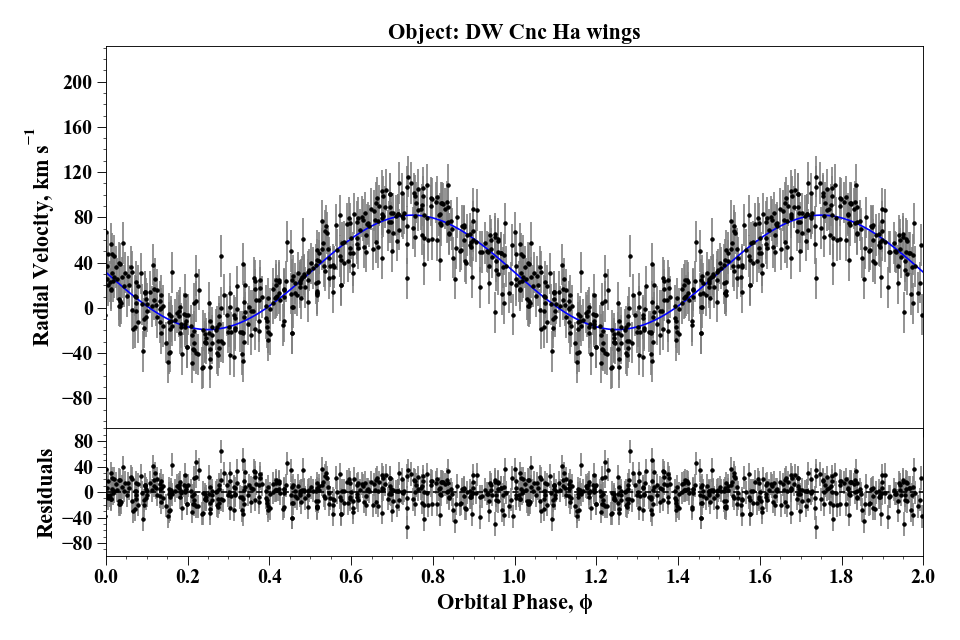}
    \caption{Radial velocity curve of the wings of the H$\alpha$ emission line for the best best solution derived from the diagnostic diagrams (see text). No evident modulation is seen in the velocity curve, or in the residuals, with the 38 min spin period. The blue line shows the best fit model. Errors have been scaled so $\chi^2_{\nu}=1$. }
    \label{fig:radial_vel1}
\end{figure}

First, we employed the two-Gaussian or {\sc gau2} option to optimize the fitting of the wings of the line, following \citet{ssh86}. The results of the diagnostic diagram are shown in Figure~\ref{fig:diag}. The best fitted values were obtained for Gaussian separations of 34 \AA ~ with individual widths of 3.5 \AA . The best orbital fit is shown in  Figure~\ref{fig:radial_vel1} (the 1$\sigma$ error bars have been scaled so that the statistical distribution $\chi^2_{\nu}=1$). This, and the subsequent fits have been obtained by running \caps{orbital}\footnote{Available at \url{https://github.com/Alymantara/orbital_fit}}
a simple least square program to determine, in general, the four orbital parameters, any of which can be set to a fixed or variable value. In our case we have set the orbital period fixed to the value mentioned above, i.e. the value calculated in Section~\ref{power-spec}. The orbital parameters are shown in the second column of Table~\ref{orbpar} (H$\alpha$ wings). Although the results confirm the orbital modulation found by Rea04 and Pea04, no convolved modulation was detected for the dwarf spin period, as that found by Pea04, (e.g. a mixed radial velocity curve like the one shown in their second panel from Top to Bottom in their Figure~7).

\begin{figure}
	\includegraphics[width=\columnwidth]{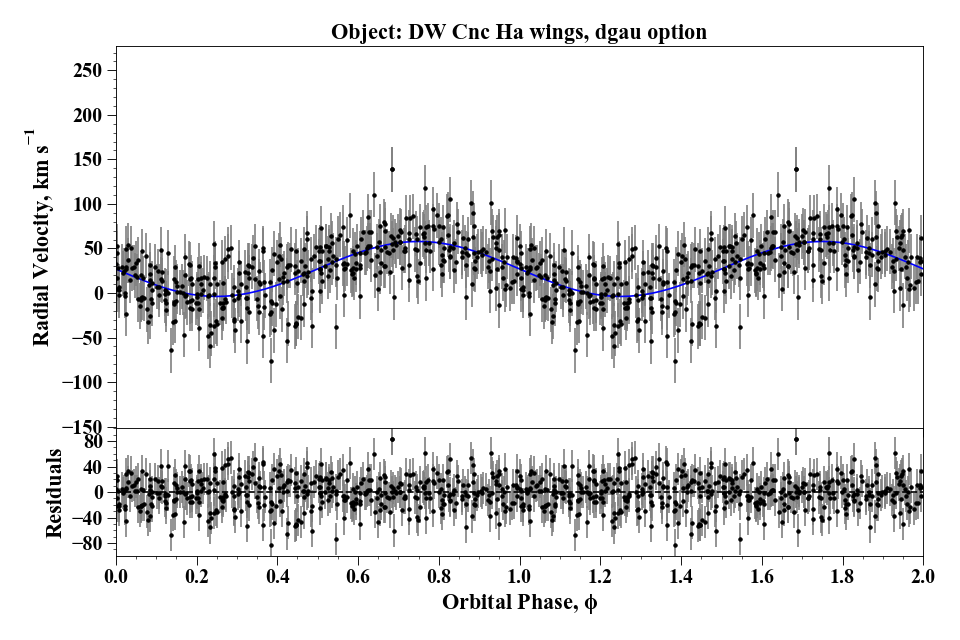}
    \caption{Radial velocity curve of the H$\alpha$ emission line, using the dgau option (see text). Again, no evident modulation is seen in the velocity curve, or in the residuals, with the 38 min spin period.  The blue line shows the best fit model. Errors have been scaled so $\chi^2_{\nu}=1$.}
    \label{fig:radial_vel2}
\end{figure}

\begin{figure}
	\includegraphics[width=\columnwidth]{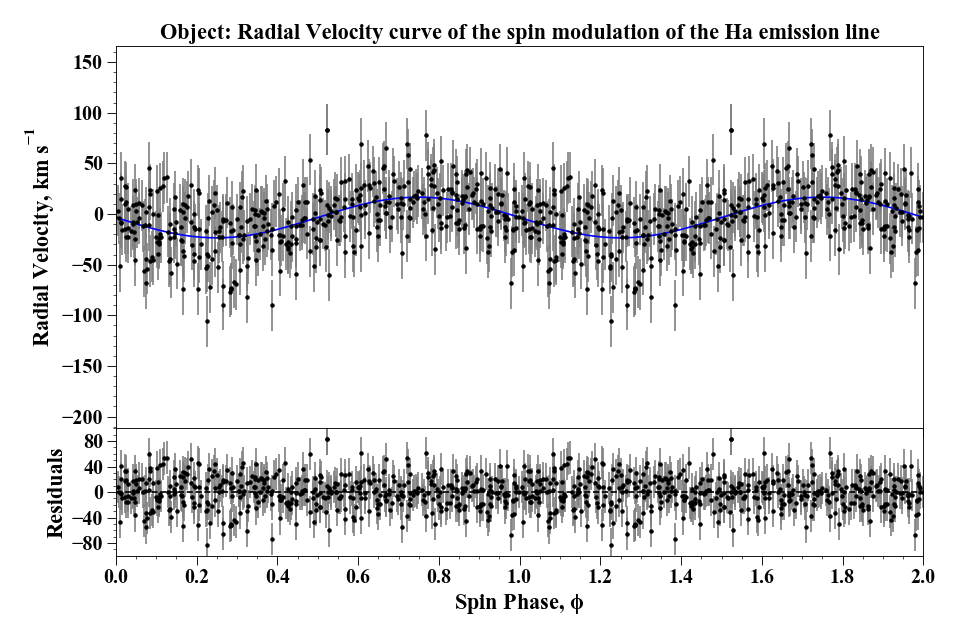}
    \caption{Radial velocity curve of the H$\alpha$ emission line, obtained after subtracting the fit of the 86 min orbital period from the data points and then folding the data with the spin period. A weak modulation is observed. }
    \label{fig:substracted}
\end{figure}

Puzzled by this lack of spin-cycle modulation,  we have convolved the H$\alpha$ emission line spectra with the derivative of a Gaussian (in our {\sc convrv} program we used the {\sc dgau} option) and used a single Gaussian, 24 \AA\  wide. This would be equivalent to the way Pea04 measured their H$\alpha$ lines. The semi-amplitude obtained using this option yielded a smaller value than that obtained from the wing-fitting method, but no modulation of the \textbf{spin period} is evident in Figure~\ref{fig:radial_vel2}.    
The calculated orbital parameters are shown in column 3~of~Table~\ref{orbpar} (H$\alpha$ dgau). The orbital parameters, $HJD_0$ and $\gamma$,  showed values consistent with the previous H$\alpha$ attempt, although the semi-amplitude yields almost half the value obtained from the H$\alpha$ wing calculation.
To make sure that our data presented no modulation from the spin period, we subtracted the best fit from the observed radial velocities. Then we folded the data with the spin period and in fact we obtain a weak modulation as can be appreciated in Figure~\ref{fig:substracted}.

\begin{table}
\centering
\caption{Orbital Parameters obtained from the H$\alpha$ line, using two different methods and from the HeI $\lambda$ 5876 \AA\ line (using the dgau method). The orbital period has been set fixed (see text).} 
\label{orbpar}
\begin{tabular}{llll}
\hline
Parameter  &  H$\alpha$ wings     &     H$\alpha$ dgau & He I dgau                \\
\hline
  $\gamma$ (km\,s$^{-1}$) & 31.4 $\pm 0.8$     & 29.5
  $\pm 1.7$  & -16.0   $\pm 1.8$   \\
   $K_1$ (km\,s$^{-1}$)   & 50.6 $\pm 1.2$     & 26.3  $\pm 1.2$  &  44.0  $\pm 2.5$   \\
  $HJD_0$*          & 0.9056 $\pm 0.0002$    & 0.9062  $\pm 0.005$ &   0.9151  $\pm 0.0005$  \\
  $P_{orb}$  (min)        & Fixed** & Fixed** & Fixed** \\
\hline
\end{tabular}
*(24559037+ days)\\
**86.10169~$\pm$~0.00031 min\\
\end{table}

\begin{figure}
	\includegraphics[width=\columnwidth]{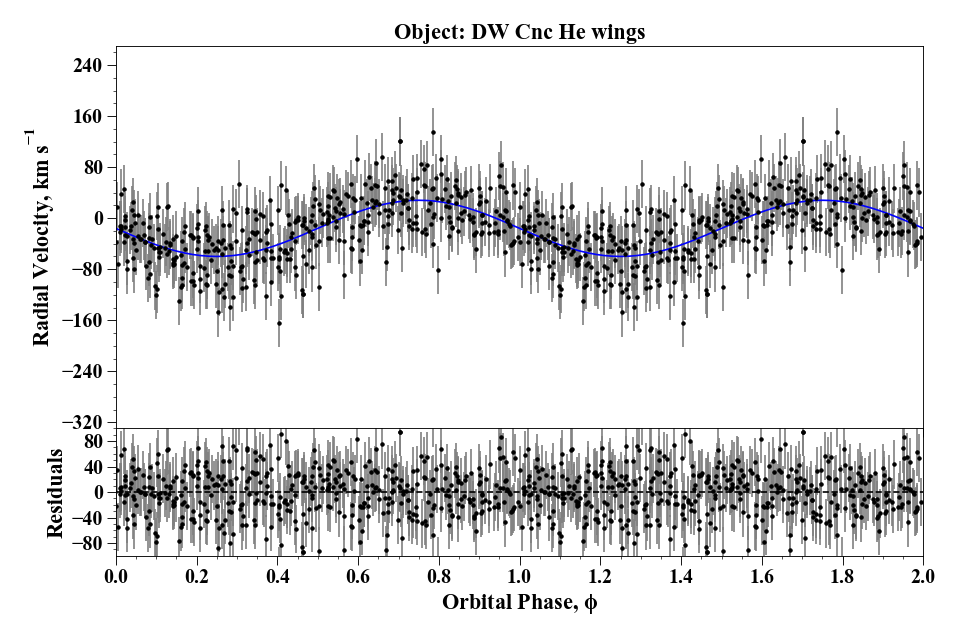}
    \caption{Radial velocity curve of the HeI $5876$ \AA\ emission line, using the dgau function.  The blue line shows the best fit model. Errors have been scaled so $\chi^2_{\nu}=1$.}
    \label{fig:radial_vel3}
\end{figure}

\begin{figure}
	\includegraphics[width=\columnwidth]{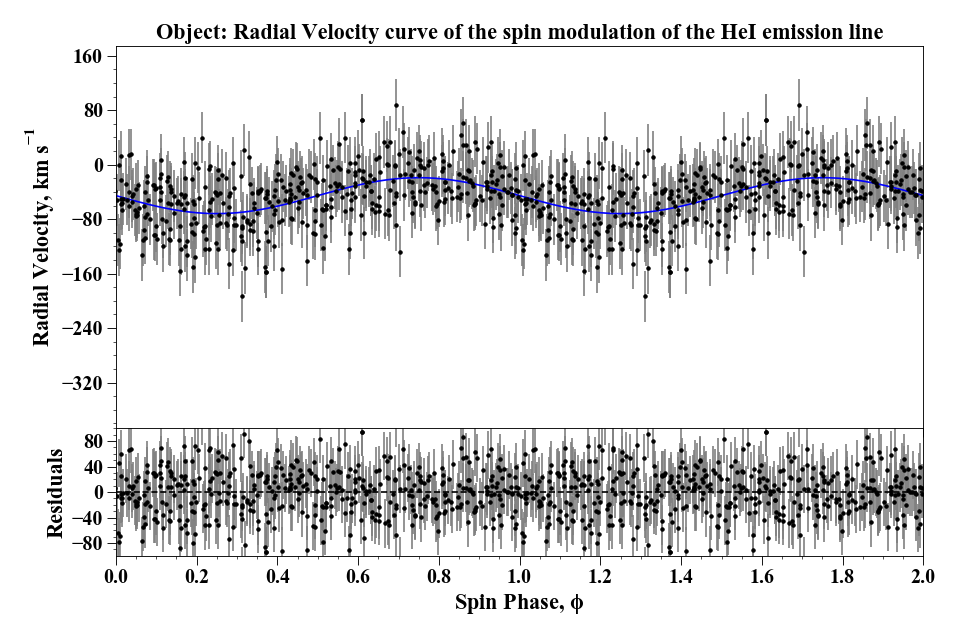}
    \caption{ Radial velocity curve of the  HeI $5876$ \AA\ emission line, obtained after subtracting the fit of the 86 min orbital period from the data points and then folding the data with the spin period. A weak modulation is observed.}
    \label{fig:substracted2}
\end{figure}

We also measured the HeI $\lambda$ 5676  \AA, using  the {\sc dgau} function in a similar manner as H$\alpha$ using a single Gaussian 23.8 \AA ~ (40 pixels) wide. The results are shown in Figure~\ref{fig:radial_vel3} and column 4 (He I dgau) of Table~\ref{orbpar}. The systemic velocity differs substantially from the H$\alpha$ results, while $K_1$ yields a value between the wing method and the single Gaussian approach. As we did with H$\alpha$ we have subtracted the best fit from the observed radial velocities and then folded the data with the spin period. Again we find a small modulation (see Figure~\ref{fig:substracted2}).

\section{Power spectrum search}
\label{power-spec}

\begin{figure}
	\includegraphics[width=\columnwidth]{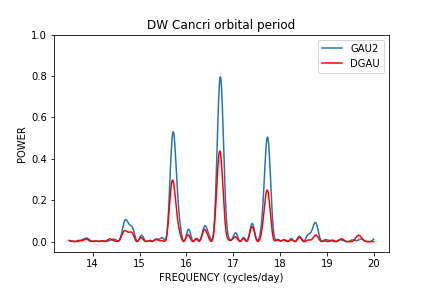}
	\includegraphics[width=\columnwidth]{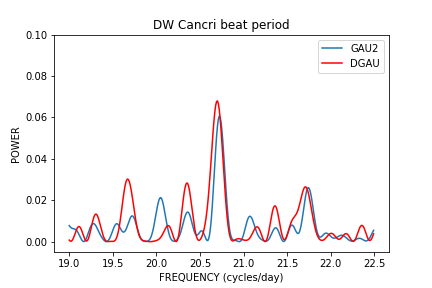}
	\includegraphics[width=\columnwidth]{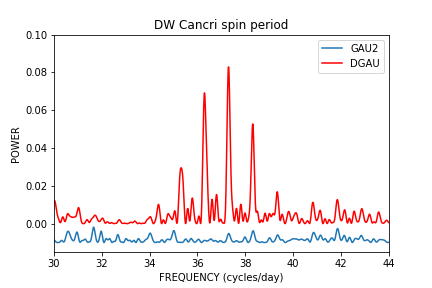}	
	\caption{Power spectrum of H$\alpha$ obtained for the {\sc dgau} and {\sc gau2} options, around the frequencies of the orbital period (upper panel), the beat-period (middle panel) and the spin period (bottom panel). In the latter we have shifted by -0.01 the {\sc gau2} result (i.e. the absence of a signal for the spin-cycle). A full explanation of the panels is discussed in the text.}
    \label{fig:poworbha}
\end{figure}

\begin{figure}
	\includegraphics[width=\columnwidth]{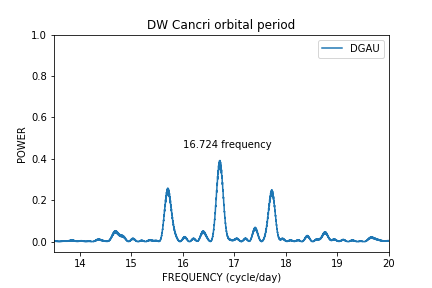}
	\includegraphics[width=\columnwidth]{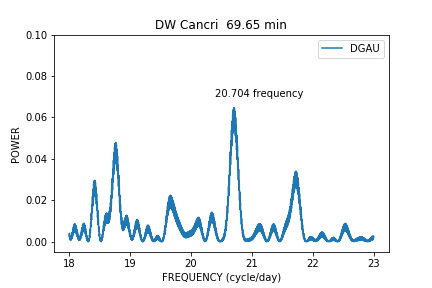}
	\includegraphics[width=\columnwidth]{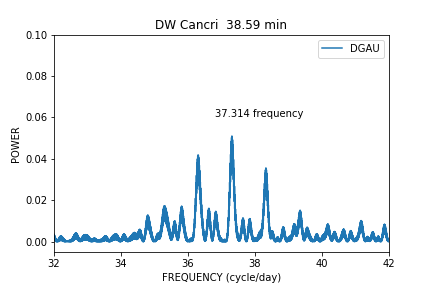}	
	\caption{Power spectrum of HeI 5876 \AA \ obtained for the {\sc dgau} option, around the frequencies of the orbital period (upper panel), the beat-period (middle panel) and the spin period (lower panel). The panels have the same scale as those in Figure~\ref{fig:poworbha}. A full explanation of the panels is discussed in the text.}
    \label{fig:poworbhe}
\end{figure}

We have made a power spectrum analysis from the measured radial velocities of H$\alpha$ and the He I~$\lambda$ 5876 \AA\ line. {\sc Lomb-Scargle}, {\sc dft} and {\sc anova} algorithms have been applied to all data using the {\sc peranso} package \citep{pav16}, including the October 2016 observations, but also using the January 2017 data only. The results in both cases are independent of the selected power search algorithms. Therefore, to keep a compact data set, we analyze hereinafter only the Lomb-Scargle results for the 2017 observations, using a {\sc python}\footnote{https://docs.astropy.org/en/stable/timeseries/lombscargle.html} package. We have first used the results of the {\sc dgau} and {\sc gau2} options for H$\alpha$. The combined results are shown in Figure~\ref{fig:poworbha}. In the upper panel we show the frequencies associated with the 86-min orbital period. In the middle panel the beat period is shown; no significant difference can be observed between the two cases. In the lower panel we show the spin period. We shifted the power spectrum of the {\sc gau2} option by -0.01 for better appreciation. No significant power signal is found using this option, while with the {\sc dgau} option a power signal is found. Note that we have scaled the upper panel from zero to one, whereas the middle and lower panels have been scaled from zero to 0.1, in order to make the comparison easier.

In the upper panel from Figure~\ref{fig:poworbha}, the orbital period is shown, for the {\sc gau2} option we found a peak at a frequency of 16.72441 $\pm$ 0.00006 cycles/d, equivalent to a period of 86.1014 $\pm$ 0.0003 min surrounded by two peaks, which are two equal aliases of 16.724 $\pm$ cycles/d. We also measured H$\alpha$ by convolving the line with the derivative of single Gaussian, 23.8 \AA\ (40 pixels) in width, using the {\sc dgau} option, the power spectrum peak is found at 16.711 $\pm$ 0.004 cycles/d, equivalent to a period of 86.17 $\pm$ 0.02 min, such being very similar to the previous result, with an almost identical adjacent structure, but with half the strength. In the middle panel (beat period) we observe, using the {\sc gau2} option a strong peak at 20.69643 $\pm$ 0.00008 cycles/d, equivalent to a period of 69.5772 $\pm$ 0.0003 min. Using the {\sc dgau} option we measured a peak with 20.7045 $\pm$ 0.005 cycles/d value, equivalent to a period of 69.650 $\pm$ 0.005 min. These peaks have a close value to that of the peak identified by Pea04 (photometrically) as the beat-period, but with a strength factor ten times smaller than the orbital period power peak. This is the first inconsistency with the results of Pea04, who find the opposite, i.e. a strong beat-period signal and a weak one for the orbital period. For the lower panel (spin period), we do not find any signal using the {\sc gau2} option, while for the {\sc dgau} option we found a weak peak present at a frequency of 37.298 $\pm$ 0.004 cycles/d, equivalent to a period of 38.608 $\pm$ 0.004 min. 

We must conclude that the wing analysis, which focuses on the inner parts of an accretion disc, or on equivalent regions at high velocities, does not detect the spin period, while regions with lower velocities do show a weak signal. This results do not resemble at all the relative strengths found by Pea04.

To complete these power searches we have done the same analysis, but now using the HeI 5876 \AA\ line using the same {\sc dgau} parameters for a single Gaussian. The results are very similar to those of H$\alpha$ using the {\sc dgau} routine. The power spectrum around the orbital period range has a peak at 16.72377 $\pm$ 0.004 cycles/d, equivalent to an orbital period of 86.105 $\pm$ 0.004 min. The beat period has a peak at 20.7045 $\pm$ 0.005 min, equivalent to a period of 69.650 $\pm$ 0.005 min. The spin period analysis yields a value of 37.314 $\pm$ 0.006 cycles/d, equivalent to a period of 38.592 $\pm$ 0.006 min. We see that for the {\sc dgau} option, the HeI power spectra are consistent with their H$\alpha$ counterparts. It is evident, from our power spectrum trials that the spin-cycle is very weak (and non existent in the H$\alpha$ wings analysis). The ratios of power signals are completely different from previous published results. 

\section{New ephemeris}
\label{ephem}

\begin{figure}
	\includegraphics[width=\columnwidth]{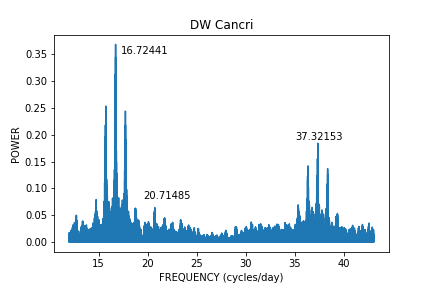}
	\includegraphics[width=\columnwidth]{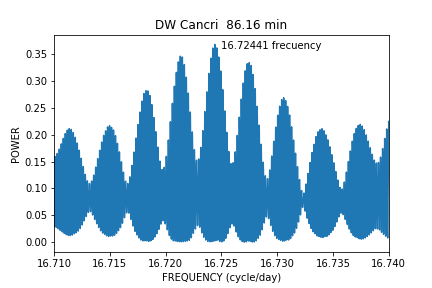}
	\includegraphics[width=\columnwidth]{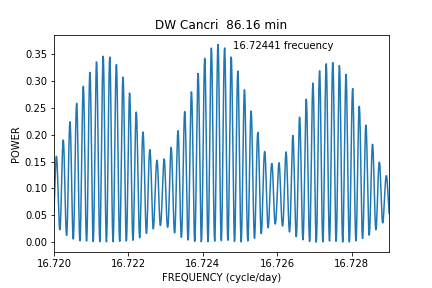}	
	\caption{Power spectrum obtained by combining the radial velocities of Pea04 with those in this paper. In the {\it upper panel} we show the dominant frequencies, while in the {\it middle and lower panel} we show a high precision search for the orbital period value (see text).}
    \label{poworb4}
\end{figure}

We now try to derive an improved orbital period value by combining the radial velocities published by Pea04 with our velocities obtained using the {\sc dgau} option, which were derived with a method equivalent to Pea04. We have included, for this calculation, the 18 spectra obtained in October 2016\footnote{The combined velocities are also found in the link mentioned in Section~\ref{sec:obs}.}. We follow the approach by \citet{hea86}: first to try to find a higher precision orbital period value around the already known orbital period value, and then use this period as a fixed value and calculate the other free parameters in a sinusoidal fit as described in Section~\ref{sec:orbper}. As expected, the combination of the data produces a mixed picture in the power spectrum domain.
Figure~\ref{poworb4} ({\it upper panel}) shows the dominant frequencies derived from this combined radial velocities. Again, as expected from this mixed data, we find a strong signal for the orbital period, while the spin period signal has a weaker strength and the beat period appears as the weakest signal. We now try to derive a new orbital period value starting at a basic 16.72 cycle/d frequency. Figure~\ref{poworb4} ({\it middle panel}) shows that the power spectrum around this frequency has a number of adjacent aliases. We suspect that, although here the 16.72441 cycle/d frequency appears unquestionable, if we increase the resolution to a step of 0.00001 cycles/d (same Figure, {\it lower panel}) we can see that each peak consists of a number of equally spaced narrow peaks. Here we find the caveat: the number of cycles that elapsed over the long time interval between Pea04's data and their own is apparently not well-constrained, which leads to closely spaced peaks in the periodogram of nearly equal strength; there is no guarantee that the strongest peak in the lower panel  is in fact the correct peak. In other words we have an uncertainty in the cycle count between the 2004 data and the present data.  Nevertheless our best result gives a value of 16.72441~$\pm$~0.00006~cycles/d, equivalent to an orbital period of 86.10169~$\pm$~0.00031 minutes. Setting this value as a fixed  parameter, we ran {\sc orbital} on the combined data to obtain the zero point. Thus, we determined a new ephemeris as:

$$ HJD = 52620.7923(5) +E 0.0597928(2).$$
\label{ephemer}

Due to the reasons described above we are aware that we are choosing an arbitrary number of cycles between the old and new data. Therefore, we will use this ephemeris only for purposes of phasing the data all together.

\section{Doppler tomography}
\label{tomo}

A time resolved spectroscopy technique, known as Doppler Tomography, based on the use of a detailed emission line profile, allow us to map the accretion flow in velocity space. A detailed formulation of this technique can be found in \citet{mah88}. We have used the H$\alpha$ and HeI $\lambda$ 5876 \AA\ lines, to produce Doppler tomograms using a newly developed \caps{PyDoppler}\footnote{Available at \url{https://github.com/Alymantara/pydoppler}} python code. This code uses the original {\sc fortran} programs, developed by Henk Spruit\footnote{Available at \url{https://wwwmpa.mpa-garching.mpg.de/~henk}}, designed to work with an {\sc idl} enviroment. 

\subsection{The H alpha Doppler tomogram}

\begin{figure}
 \centering
	\includegraphics[width=\columnwidth]{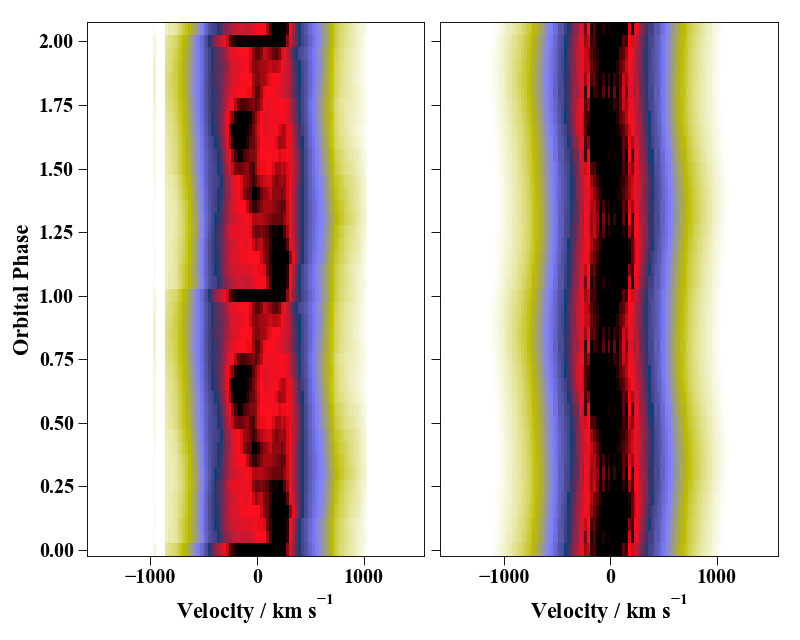}
	\includegraphics[width=\columnwidth]{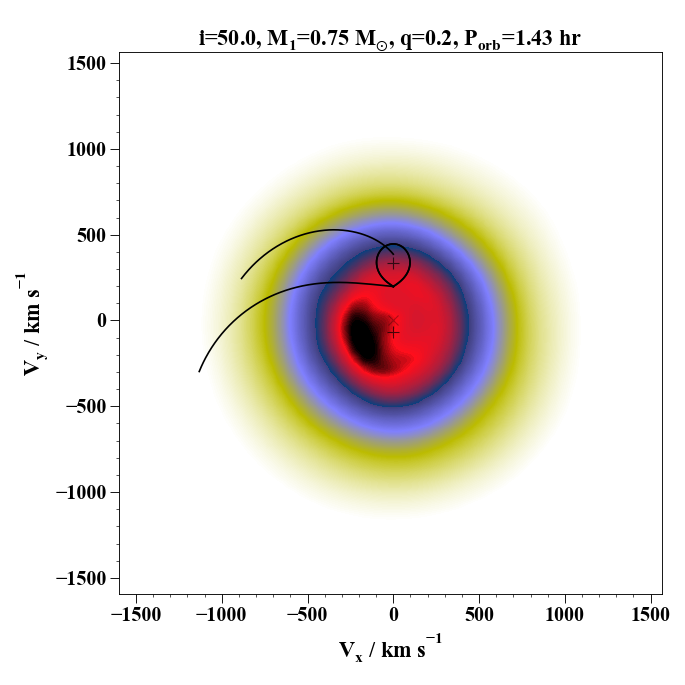}
    \caption{Trail spectra and Doppler Tomogram spectra of the He I~$\lambda$ 5876~\AA\ emission line. The relative emission intensity is shown in a scale of colors, where the strongest intensity is represented by black, followed by red, then blue, and finally yellow which depicts the weakest intensity. The color white represents the lack of emission.}
    \label{fig:reco-ha58}
\end{figure}

We have constructed a Doppler tomogram based on the emission line profile of the H$\alpha$ emission line. The observed trail-spectrum is shown in the {\it upper left} panel ({\sc spec}) of Figure~\ref{fig:reco-ha58}, while the reconstructed one is depicted in the {\it upper right} panel ({\sc reco}). The Doppler tomogram is shown at the {\it bottom} ({\sc stream}) of the same Figure. Both the {\sc spec} and {\sc reco} show a clear single and broad sine-wave with the orbital phase, but shifted such that the maximum velocity appears around phase 0.15 and not 0.25. It also shows a clear intensity from low to higher velocities. The tomogram, reveals a single emission area, centered at velocity coordinates $V_x$ = -200 km\,s$^{-1}$; $V_y$ = -25 km\,s$^{-1}$ a region where SW Sex star usually show their emission \citep[e.g][]{hel96}. 

\subsection{The He I Doppler tomogram}
\label{hei-tomo}

The simple sinusoidal behaviour of the HeI $\lambda$5876 \AA\ line observed in the {\sc spec} trail spectra in not as clear as that observed in H$\alpha$. A more complex behaviour is detected. In fact, the tomogram reveals an emission distribution with a weak spiral-arm-like structure. This spiral structure has been the landmark of IP~Peg, whose spiral arms were detected by \citet {shh97} at the end of a rise and close to maximum light. These authors explain the spiral structure in IP~Peg as the consequence of a large accretion disc on which the secondary star induces tidal waves in the outer disc. 

\subsection{General remarks about the doppler tomograms}
\label{gen-tomo}
The results in our doppler tomography show, from both lines, that there was no accretion disc formed at the time of our observations. The H$\alpha$ tomogram shows that the emission was accumulated near a velocity zone usually seen in SW~Sex stars. However, this object does not show the usual characteristic features of SW~Sex stars \citep[e.g.][and references therein]{hea03, hel96}. The He I~$\lambda 5876$~\AA\ tomogram reveals a weak spiral-arm structure. Although this is the landmark of the IP~Peg-like systems (see Subsection~\ref{hei-tomo}), \dw ~was not observed near outburst. Its structure seems, more likely, to reflect the possibility that some of the He~I rich material might still be flowing into the magnetic pole. Whether or not these arms will be developed into full spiral-arms, like those observed in DQ~Her by \citet{bea10} during quiescence, after \dw returns to its high-state, will have to wait for new observations.

\section{Discussion}
\label{sec:discussion}

We have found DW~Cnc in a state where the spin period signal is very weak compared with that of the orbital period, in stark contrast with the results of Rea04 and Pea04, who find a strong spin period signal greater than, or at least equal to, that of the orbital period modulation. The emission from a magnetic pole in an asynchronous rotating white dwarf is the \textit{ essential credential of an intermediate polar} as put succinctly by Pea04. If this is the case in DW~Cnc, one has to ask the question: what changed between the time of the observations around 1999-2004 and 2017-2018?

Such a change could be due to a substantial diminishing of the mass accumulated in the external regions of the disc precluding the accretion flow into the magnetic poles of the white dwarf and a subsequent emission from them. If DW~Cnc is in fact a VY~Scl system this would provide us with an  explanation. VY~Scl are cataclysmic variables whose light curves are characterized by occasional drops from steady high states into low states lasting up to several hundred days \citep[see][and references therein]{kac98,lea99}. As pointed out by \citet{lea99}, these low states probably result from episodes of low mass transfer from the companion star. \citet{kac98} point out that the most likely cause of the low mass transfer in these timescale is the presence of star-spots, a model initially suggested by \citet{lap94} to show the apparent absence of dwarf novae in the period gap. Is \dw indeed a VY~Scl system as suggested by Rea04? (see their Section 3.1). 

The {\sc aavso} observational data, which spans almost 20 years, presented in Section~\ref{intro}, shows us that \dw has spent most of this lapse at a mean value of $\sim$ 15.5 mag. Scarce observations show that the system was at one magnitude brighter before 2004, with an occasional short outburst in 2007 (and possibly two more in 2008 and 2009). There is a sudden drop to a low-state at the beginning of 2018 and throughout 2019, with a recovery to 15.5 mag at the beginning of 2020. The above description of the {\sc aavso} light curve of \dw shows the true markings of a  VY~Scl system. The low-state observed in 2018 and 2019, suggests that indeed \dw was observed at an episode of low mass transfer from the companion. The weak signal of the spin period that we obtained can be explained if the outer disc in \dw was being replenished; the outer disc was still in the process of gaining mass and no substantial mass transfer occurred at the moment of our observations. If we are correct, we should see \dw in the near future, functioning again as a full VY~Scl intermediate polar. This process might be starting to occur, since the magnitude in early 2020 appeared to have reached its mean value again. This process might be similar to that of FO~Aqr \cite[see][and references therein]{hal17,lea16} which has shown a low-state only once (in 2016 and recovered slowly to its high-state in a matter of months). It is worth mentioning that \dw is the shortest orbital VY Scl system, its closest relative being V442~Oph, which has an orbital period of 2.98 hours \citep{gaz88}. For an orbital period distribution of VY~Scl \citep[see][and references therein]{ver97,han98}. 

\begin{figure}
 \centering
	\includegraphics[width=\columnwidth]{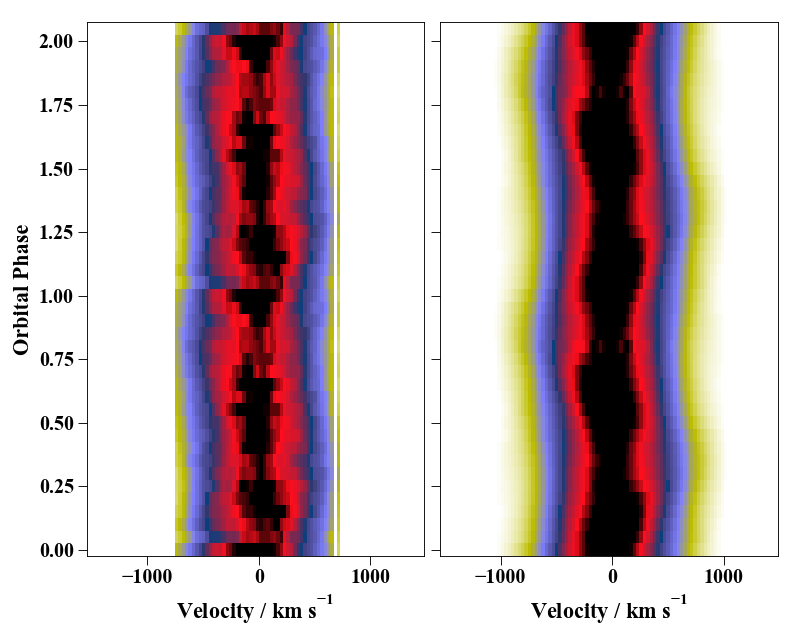}
	\includegraphics[width=\columnwidth]{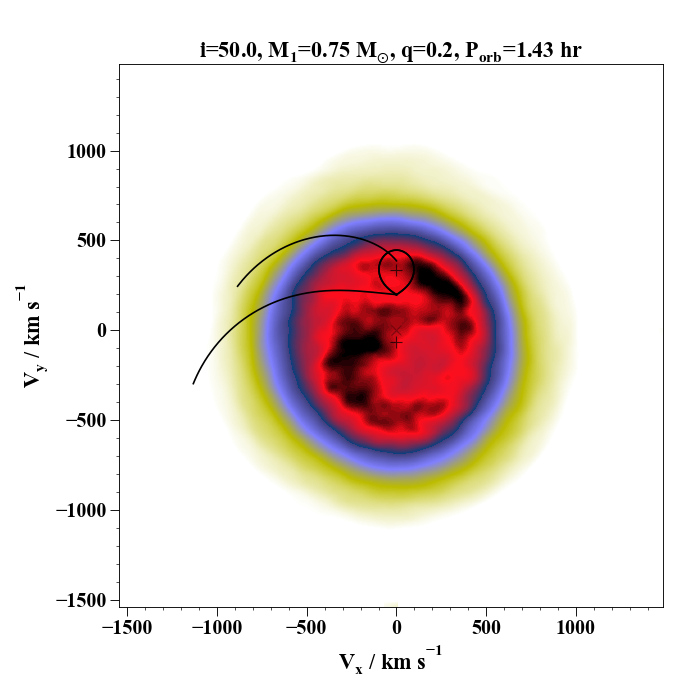}
    \caption{Trail spectra and Doppler Tomogram spectra of the He I~$\lambda$ 5876~\AA\ emission line. The relative emission intensity is shown in a scale of colors, where the strongest intensity is represented by black, followed by red, then blue, and finally yellow which depicts the weakest intensity. The color white represents the lack of emission.}
    \label{fig:reco-he}
\end{figure}

We also explore as a possible explanation to this change, the 2007 outburst event observed by \citet{cea08}. \citet{hal17} show that dwarf novae outbursts in intermediate polar systems must have low mass transfer rates and large magnetic fields in order to keep the disc stable, i.e. to maintain them on the cold equilibrium branch of the well accepted thermal-viscous disc instability model. However, there is a caveat to this scenario, as follow up observations by \citet{cea08} still show the asynchronous 38.6 min spin-period signal of the WD, as well as a signal at 73.73 min, similar to that found by \citet{uea02}, as mentioned in Section~\ref{intro}. Furthermore, the X-Ray findings by \citet{nea19} obtained in 2012 also show, within their errors, the spin-period. The 2007 outburst does not seem as a likely explanation for the change in behavior of DW~Cnc, although we do not rule it out entirely.  

\section{Conclusions}
\label{sec:conclusions}

Spectroscopic observations of \dw carried out in October 2017 and mostly within a week in January 2018 show that this intermediate polar has almost  ceased to show the 38 min spin cycle emission arising from the asynchronous magnetic white dwarf, in stark contrast with the observations by \citet{rea04} and \citet{pea04} made during 1999-2004.

We have discussed two possible scenarios to explain this change. The first and more convincing one is that \dw is indeed a magnetic VY~Scl-type system as proposed by \citet{rea04}. This is confirmed by our review of the {\sc aavso} light curve of \dw  done in Section~\ref{sec:discussion} and in the fact that our observations were made during a low-state. As discussed before, this could result in an episode of low mass transfer rate from the companion star \citep{lea99} which could be responsible for the lack of the spin-cycle signal. 
If this scenario is correct, we predict that \dw will recover its intermediate polar characteristics at the beginning of 2020, as it has already returned to its high-state brightness.

The second scenario proposes that the weak spin cycle signal is the consequence of a short outburst detected in 2007 by \citet{cea08}, which could cause the outer accretion disc to be depleted and therefore unable to transfer enough material into one of the magnetic poles. However this scenario is not consistent with the fact that the 38 min signal is still present after the outburst nor with the consideration that the same signal is present in the X-Rays found by \citet{nea19}.

We have combined Pea04 radial velocities with ours to improve the base-line and to obtain a more precise orbital period determination. We did a Lomb-Scargle search and found a value of 86.10169~$\pm$~0.00031 min for the orbital period, 69.650~$\pm$0.005 min for the beat period and a weak signal at 38.608$\pm$0.004 min for the spin period.  The strength of the signals that we found are not consistent with those from Pea04, since their orbital period signal is weaker than the signal found for their beat period, whereas we found the opposite.

The Doppler Tomography shows a concentration of H$\alpha$ emission similar to those found in SW~Sex stars, centered at velocity coordinates $V_x$ = -200 km\,s$^{-1}$; $V_y$ = -25 km\,s$^{-1}$, although \dw does not present the usual characteristics of the SW~Sex-type stars. The He I~$\lambda$ 5876 \AA\ emission shows a weak spiral-arm which suggests that some material might still be flowing into the magnetic pole. As pointed out in the discussion, whether or not these arms will develop into full spiral-arms, after \dw returns into its high-state, will have to be confirmed with new observations.

\section*{Acknowledgements}

The authors are indebted to DGAPA (Universidad Nacional Aut\'onoma de M\'exico) for financial support, PAPIIT project IN114917.

\bsp
\label{lastpage}
\end{document}